\documentclass[prb,reprint,superscriptaddress,amsmath,amssymb,a4paper,aps]{revtex4-1}

\usepackage[dvipdfmx]{graphicx}

\bibpunct{[}{]}{,}{n}{}{}

\usepackage{newtxtext,newtxmath}
\usepackage{comment}
\usepackage[dvipdfmx,colorlinks=true,linkcolor=black,urlcolor=black,citecolor=black]{hyperref}

\newcommand{\eqn}[1]{
\begin{eqnarray}
	#1
\end{eqnarray}
}

\newcommand{\siotwo}{\rm{SiO_2}}

\begin{document}

\title{Crystal Structure Prediction Supported by Incomplete Experimental Data}
\author{Naoto Tsujimoto}
\affiliation{Department of Physics, University of Tokyo, 7-3-1 Hongo, Bunkyo-ku, Tokyo 113-0033, Japan}
\author{Daiki Adachi}
\affiliation{Department of Physics, University of Tokyo, 7-3-1 Hongo, Bunkyo-ku, Tokyo
113-0033, Japan}
\author{Ryosuke Akashi}
\affiliation{Department of Physics, University of Tokyo, 7-3-1 Hongo, Bunkyo-ku, Tokyo
113-0033, Japan}
\author{Synge Todo}
\affiliation{Department of Physics, University of Tokyo, 7-3-1 Hongo, Bunkyo-ku, Tokyo
113-0033, Japan}
\affiliation{Institute for Solid State Physics, University of Tokyo, 5-1-5 Kashiwanoha, Kashiwa, Chiba 277-8581, Japan
}
\affiliation{Research and Services Division of Materials Data and Integrated System, National Institute for Materials Science, 1-2-1 Sengen, Tsukuba, Ibaraki 305-0047, Japan
}
\author{Shinji Tsuneyuki}
\affiliation{Department of Physics, University of Tokyo, 7-3-1 Hongo, Bunkyo-ku, Tokyo
113-0033, Japan}
\affiliation{Institute for Solid State Physics, University of Tokyo, 5-1-5 Kashiwanoha, Kashiwa, Chiba 277-8581, Japan
}

\date{\today}

\begin{abstract}
The prediction of material structure from chemical composition has been a long-standing challenge in natural science. Although there have been various methodological developments and successes with computer simulations\cite{CJP11,AL02,SK83,JP90,YW10,YW12,ARO06,CWG06}, the prediction of crystal structures comprising more than several tens of atoms in the unit cell still remains difficult due to the many degrees of freedom, which increase exponentially with the number of atoms. Here we show that when some experimental data is available, even if it is totally insufficient for conventional structure analysis, it can be utilized to support and substantially accelerate structure simulation. In particular, we formulate a cost function based on a weighted sum of interatomic potential energies and a penalty function referred to as ``crystallinity'', which is defined using limited X-ray diffraction data. This method is applied to well-known polymorphs of $\rm{SiO_2}$ with up to 96 atoms in the simulation cell to find that it reproduces the correct structures efficiently with a very limited number of diffraction peaks. The penalty function is confirmed to destabilize the local minima of the potential energy surface, which facilitates finding the correct structure. This method opens a new avenue for determining and predicting structures that are difficult to determine by conventional methods, such as surface, interface, glass, and amorphous structures. 
\end{abstract}
\nopagebreak
\maketitle
\nopagebreak
\clearpage
Since the invention of the X-ray diffraction method about a century ago, various diffraction techniques have been developed to determine the atomistic structure of objects, and they have served as essential tools over a broad range of research fields, such as solid state physics and biochemistry. However, structure determination becomes unreliable if the experimental diffraction data is incomplete, for example when the diffraction intensity is weak, the resolution is low, or there is a large amount of background noise. This weak point is sometimes a serious obstacle for modern materials science. 
The recent trend respects pursuit of novel functionalities of materials in nanoscale and/or under extreme condition such as high pressure and temperature. However, it is mostly impossible to obtain evident experimental data for such systems.

A way unaffected by these experimental complexities is theoretical structure prediction\cite{SMW08}. The modeled interatomic potential or first-principles electronic structure calculations based on density functional theory are used to calculate the total energy of the system $E$, as a function of atomic configuration. Using the chemical composition of the entire system as an input enables realistic and energetically stable structures to be derived through the optimization of $E$. 
Thanks to recent development of various efficient optimization methods such as random sampling\cite{CJP11}, metadynamics\cite{AL02}, simulated annealing\cite{SK83,JP90}, particle-swarm optimization\cite{YW10,YW12} and genetic algorithms\cite{ARO06,CWG06}, theory-based methods are rapidly becoming practical, even obtaining predictive power.
A highly successful case was a study on hydrogen sulfide under pressure. A high critical temperature ($T_c$) superconducting phase was predicted\cite{YL14,DD14} prior to the experimental observation\cite{APD15}, and the structure of the phase in a pressure cell was later confirmed by comparison of the experimentally observed and predicted diffraction data\cite{EM16}. However, the applicability of such theoretical schemes is also severely limited by the complexity of the target material. As a system becomes more complex, the number of trivial minima in the multidimensional energy landscape grows significantly and the computational cost of reaching the relevant minima increases.
A breakthrough is thus required to meet the needs in the modern materials science.

\begin{figure}
\includegraphics[width=90mm]{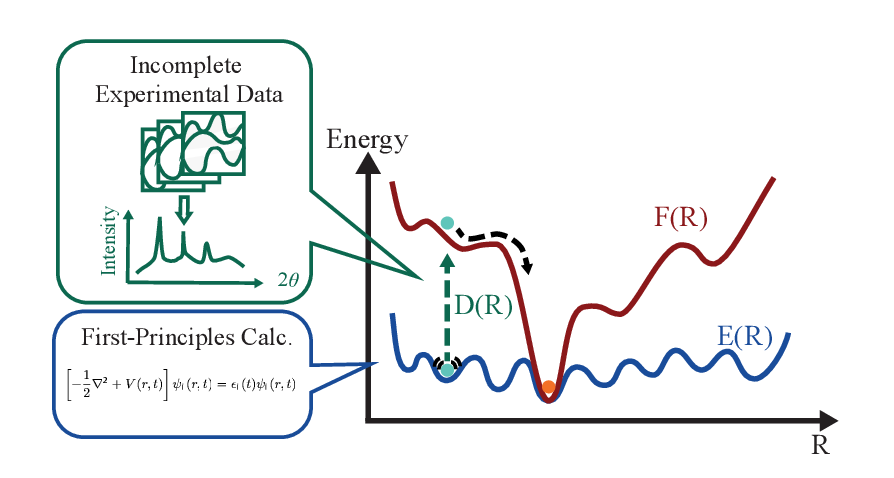} 
\caption{\label{fig1}
{ Schematic image of method of utilizing available experimental data for crystal structure prediction.}  The aim of crystal structure prediction is to find the global minimum (orange sphere) of a potential energy surface $E(\bf R)$ (blue line), where $\bf R$ denotes atomic coordinates. Although accurate calculation of the potential energy is now available, finding the lowest energy is still difficult because the search space $\bf R$ is high dimensional and the potential energy surface $E$ has many local minima (light blue sphere). To fill the trivial local minima, a penalty function $D$, defined with available experimental data is introduced. We expect that it is easier to find the global minimum of the cost function $F$ (red line), which is the sum of $E$ and $D$, than the optimization using only $E$.}
\end{figure}

Here, we propose a new direction of theoretical development for the structure determination problem that paves the way to a systematic improvement in the ability to reach stable structures. The central idea is to utilize incomplete experimental data to improve the efficiency of energy optimization, i.e., joint optimization of the total energy and auxiliary functions implemented with experimental data is performed instead. Such an approach has been pursued before in the experimental community. Direct space methods\cite{CR07} and data-assisted structure searches\cite{PG17} have been used to optimize the atomic configuration so that experimental diffraction data is reproduced. There have been attempts to use an additional energy cost function to support efficient fitting; however, this has included the assumption that the observed data is sufficiently accurate\cite{,HP99,OJL00,GWT01}. Instead, we develop an inverse approach in which the theoretical optimization is supported by partial experimental data. The cost function is represented as:
\eqn{\label{F}
F({\bf{R}}) = E({\bf{R}}) +  \alpha N  D[g({\bf{R}}), g_{\rm{obs}}],
}
where $E$ is the potential energy calculated for atomic position $\bf{R}$, $g$ denotes an observable function of $\bf{R}$, $D$ denotes the penalty function that represents `distance'  or the metric functional, which becomes minimum when $g = g_{\rm{obs}}$, $\alpha$ is a control parameter, and $N$ is the number of atoms.
The correct structure that reproduces the experimental data remains stable, while other experimentally irrelevant structures become unstable.  It thus becomes easy to predict the correct structure, as illustrated in Fig.~\ref{fig1}. The value of $\alpha$ is tuned so that the energy barrier in $E$ is overcome (see Methods for details). Any of the optimization methods mentioned above can be applied to locate the global minimum of the cost function $F$. 

In this work, we demonstrate an efficient implementation of a structural search with $F$, assuming an example situation where observed X-ray diffraction data is incomplete. A part of the X-ray powder diffraction data is used to formulate the penalty function $D$:  
 \eqn{\label{D}
 D({\bf{R}}) &=&1 -\lambda({\bf{R}}),
}
where $\lambda$ denotes the ``crystallinity'' defined using experimental data for a specific crystal as 
\eqn{\label{lambda}
\lambda &=& \frac{\displaystyle \sum_{ \theta_{\rm{obs}}} \int_{\theta_{\rm{obs}-\Delta}}^{\theta_{\rm{obs}+\Delta}} I_{\rm{calc}}(\theta)d\theta}{\int I_{\rm{calc}}(\theta)d\theta}.
}
Here, $ I_{\rm{calc}}(\theta)$ denotes the intensity of the calculated diffraction pattern, $\theta_{\rm{obs}}$ denotes the peak positions in the experimentally observed (referenced) diffraction pattern and $\Delta$ denotes the diffraction angle resolution. In the numerator, the intensities of calculated diffraction patterns, $I_{\rm{calc}}$, are summed up only at the observed peak positions, $\theta_{\rm{obs}}$. The important point is that the intensities of the observed diffraction peaks are not considered, but only the peak positions, which are insufficient to determine the structure by themselves, are used.
Therefore, intensity modifications due to the texture of the sample, i.e., the preferred orientation effect, do not need to be corrected.
The crystallinity $\lambda$ equals 1 when the structure is correct.  

\begin{figure}
\includegraphics[width=75mm]{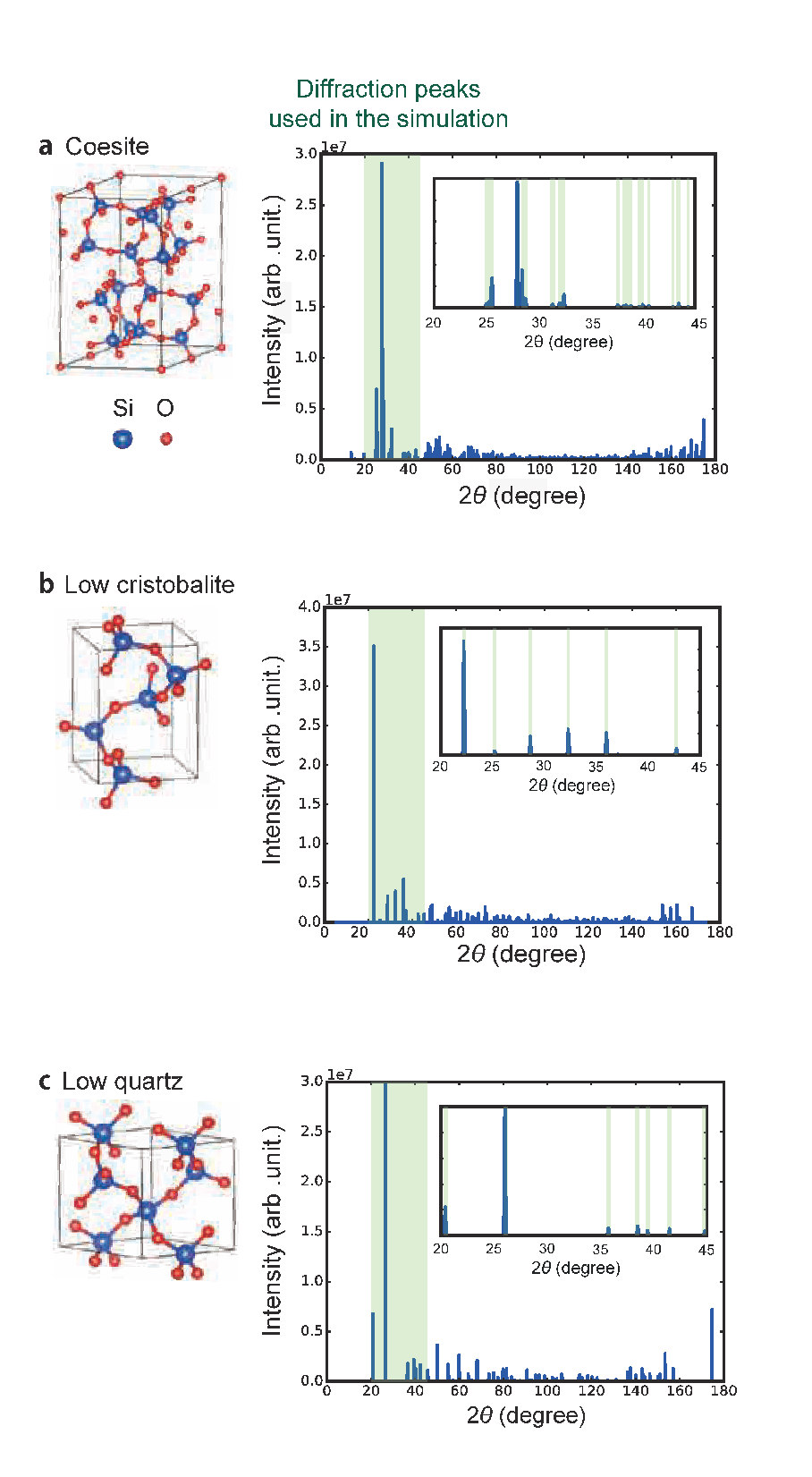} 
\caption{\label{fig2}
{ Crystal structures of three polymorphs of $\rm{SiO_2}$ and the corresponding powder X-ray diffraction patterns.}  Crystallographic unit cells and diffraction patterns for {\bf a,} coesite, {\bf b,} low cristobalite, and {\bf c,} low quartz. Si and O atoms are represented by blue and red spheres, respectively. Only information of peak positions from $20^\circ$ to $45^\circ$ (the green squares) is used to define the penalty function $D$.   }
\end{figure}

This method was applied to the well-known polymorphs of $\siotwo$. 
The potential energy surface for $\siotwo$ has numerous local minima; therefore, it is difficult to reach the experimentally observed crystal structures based only on knowledge about the chemical composition. Thus, it is a good material to demonstrate the effectiveness of the proposed method.
Three polymorphs of $\siotwo$ were adopted as target structures: coesite, low cristobalite and low quartz, among which the energy differences are small. The correct structures and their powder X-ray diffraction patterns are shown in Fig.~\ref{fig2}. The simulation cells are a $2\times1\times1$ supercell containing 96 atoms for coesite, a $2\times2\times2$ supercell containing 96 atoms for low cristobalite, and a $2\times2\times2$ supercell containing 72 atoms for low quartz. Each penalty function $D$ is formulated based only on peak positions from $20^\circ$ to $45^\circ$, as shown in the insets in Fig.~\ref{fig2}. For simplicity, the cell parameters are fixed (see Supplementary Information). Simulated annealing for molecular dynamics\cite{SK83,JP90} was used for global optimization (see Methods for details).

\begin{figure*}[t]
\includegraphics[width=120mm]{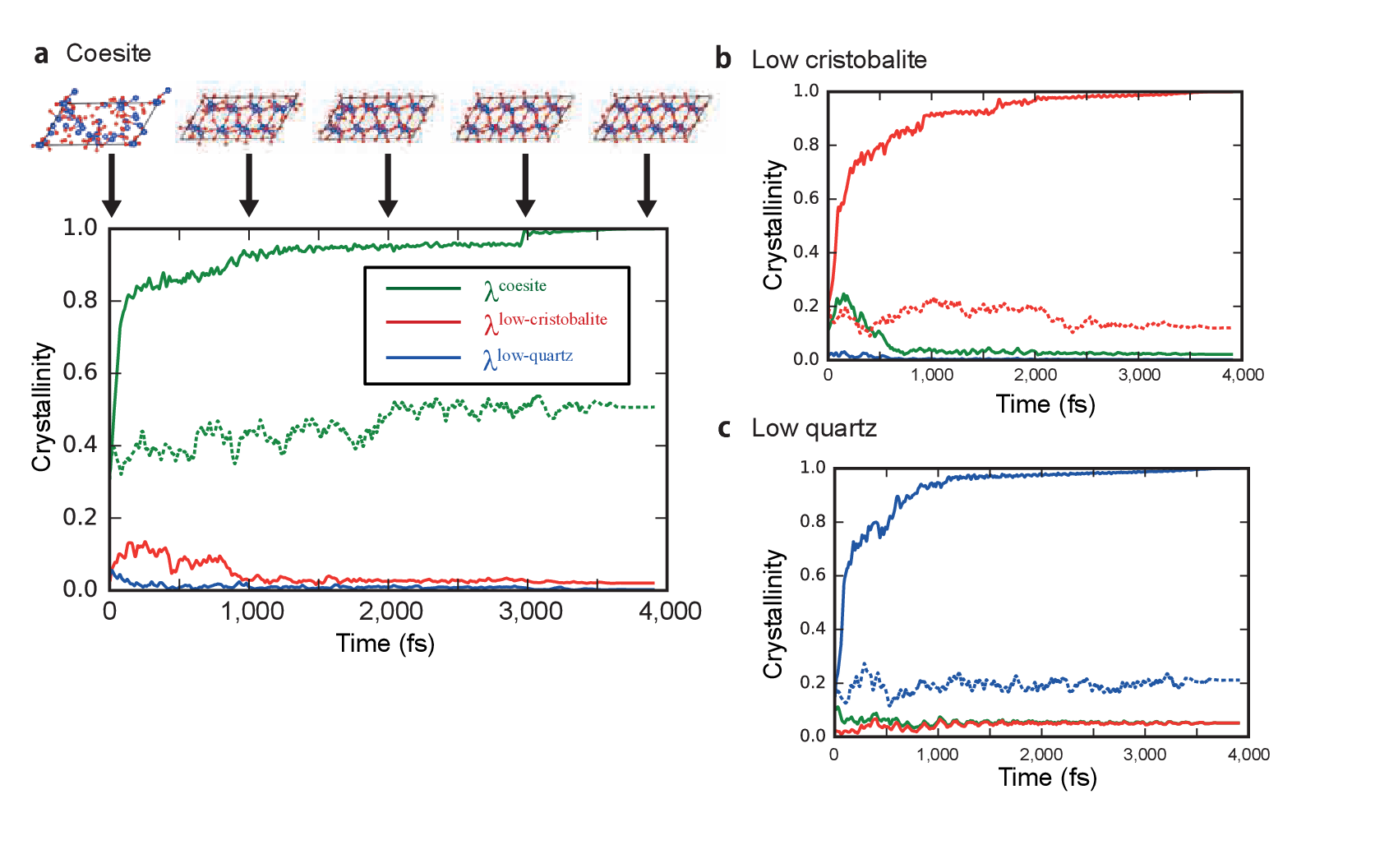} 
\caption{\label{fig3}
{ Finding correct structure supported by incomplete X-ray diffraction patterns.}  {\bf a,} The penalty function $D^{\rm{coesite}}$ is defined using the diffraction data for coesite. The green (red, blue) solid line represents the time evolution of crystallinity for coesite (low cristobalite, low quartz) during optimization of the cost function. The green dashed line shows the time evolution of crystallinity for coesite during optimization using only the potential energy.  {\bf b, c}  The penalty function is defined using the diffraction data for low cristobalite and low quartz.  The parameter $\alpha$ is set to $\frac{4\alpha}{9k_B} = 20,000 \,\rm{K}$ in all cases. }
\end{figure*}

To compare the efficiency of the scheme with $F$ and $E$, a large number of structure search trials were conducted with random atomic configurations as inputs. 
A typical method (simulated annealing with potential energy $E$) failed to find the target structure out of 500 samples, which reflects the complexity of the $\siotwo$ potential surface.
In contrast, with $F$, the target structures were correctly reached with high efficiency; 12, 109 and 148 samples out of 500 reached the global minimum for coesite, low cristobalite and low quartz, respectively. Fig.~\ref{fig3} shows the time evolution of ``crystallinity'' during the simulation for typical trials. When $F$ is formulated with the ``crystallinity'' from the diffraction data for coesite (panel a), the corresponding structure is correctly reproduced, as represented by the ``crystallinity'' (solid line) reaching 1.0. Similar results also apply when $F$ is defined by the ``crystallinity'' with the peak positions of low cristobalite and low quartz (panels b and c). Note that the ``crystallinity'' is kept far smaller than 1.0 by a typical simulation with $E$ (dashed lines), which indicates that the configuration $R$ is trapped around experimentally irrelevant energy minima.

\begin{figure*}
\includegraphics[width=140mm]{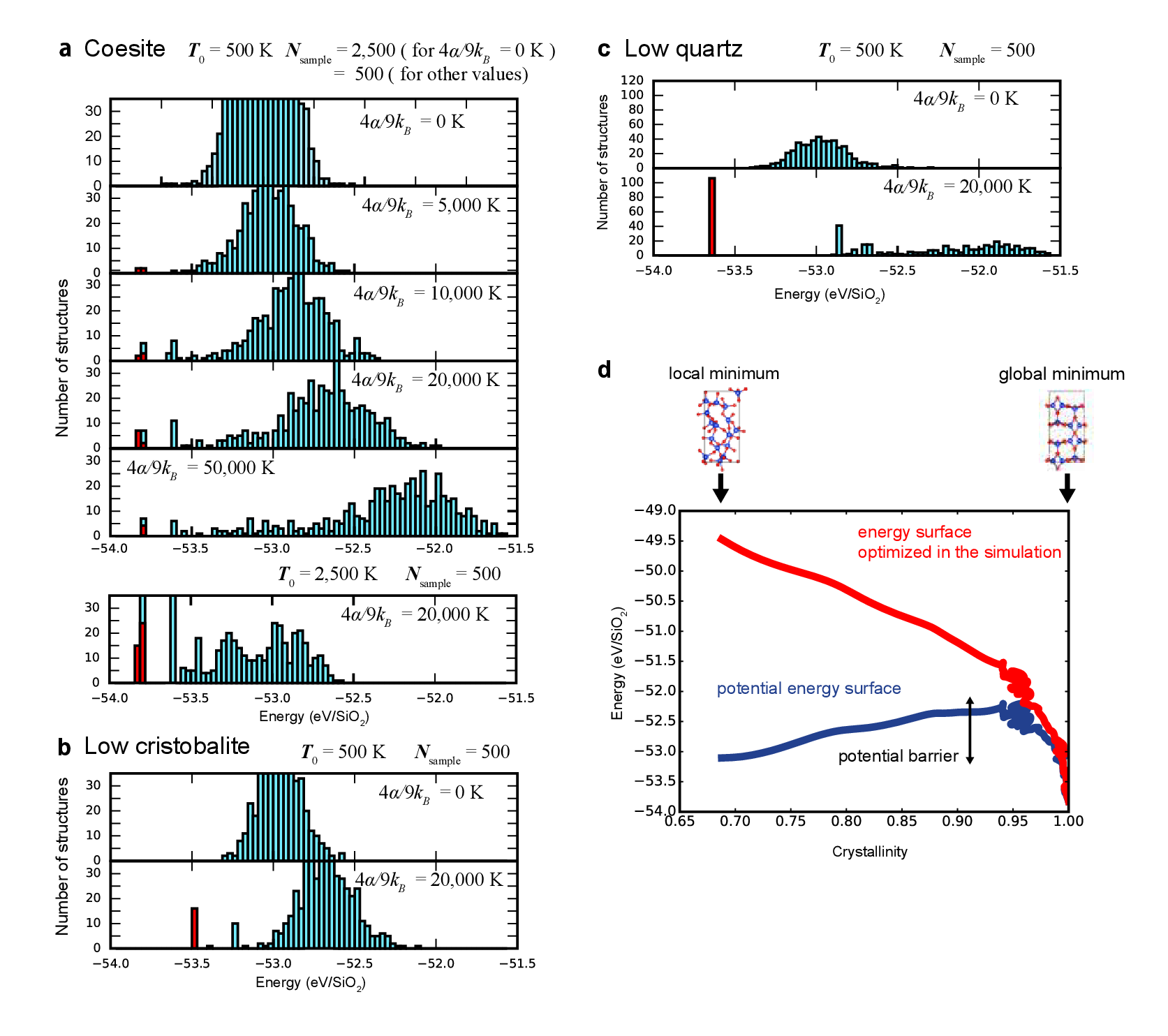} 
\caption{\label{fig4}
{ $\alpha$ dependence of the potential energy $E$ distributions for optimized structures and change in $E$ and the cost function $F$ during successful optimization of $F$. }  {\bf a,} The target material is the unit cell of coesite, which contains 48 atoms. 2,500 (500) samples were optimized for $\frac{4\alpha}{9k_B} = 0.0\, \rm{K}$ (other values) by simulated annealing at low temperature (500 K).  The most stable structure corresponds to the red bars. Simulated annealing at high temperature (20,000 K) was also performed (lower panel). {\bf b,} The target material is a $2\times2\times1$ supercell of low cristobalite, which contains 48 atoms. {\bf c,} The target material is a $2\times2\times1$ supercell of low quartz, which contains 36 atoms.  {\bf d,} Starting from the metastable structure shown in the upper left, the correct structure of coesite as shown in the upper right is found by optimization of the cost function, as depicted by the red line. The potential energy $E$ is extracted from the cost function $F$, as shown by the blue line. The potential barrier between the two structures, which makes it difficult to locate the global minimum, disappears with the penalty function. 
$\frac{4\alpha}{9k_B} = 20,000 \,\rm{K}$ in this simulation. }
\end{figure*}

Here we show that the structure of the hypersurface of the function $F$ is significantly modified from $E$ by increasing the mixing factor $\alpha$, which leads to a substantial increase in the probability of determining the correct structures. Fig.~\ref{fig4} summarizes the potential energies for the final configuration reached by the optimization of $F$ by simulated annealing at low temperature ($500$ K). None of the three target structures was reached within 2,500 (500) trials for coesite (low cristobalite and low quartz) by the optimization with $\frac{4\alpha}{9k_B} = 0.0\,\rm{K}$ (Fig.~\ref{fig4}a--c, uppermost panels). In contrast, the number of successful trials increases with increasing $\alpha$ (Fig.~\ref{fig4}a--c, red bars). Note that the success probability decreases when $\frac{4\alpha}{9k_B}$ is too large, which corresponds to the neglect of the potential energy. The optimum value of the factor $\frac{4\alpha}{9k_B}$ is in the order of the melting temperature, which is plausible to overcome the energy minima. By changing the annealing temperature $T_0$, the success rate can be further improved (panel a, lower). The upward shift of the final energy distributions with increasing $\alpha$ indicates that escape from the energy minima is generally facilitated at the expence of the increase in the median of the energy. 
Panel d of Fig.~\ref{fig4} represents the trajectories of $E$ and $F$ during optimization with $F$ from an experimentally irrelevant metastable structure to the target structure. It is clear that the minimum for the $E$ surface corresponds to the middle of the slope for the $F$ surface, and the basin of energy minimum for $E$ is buried. Therefore, optimization with the $F$ surface easily yields the target structure. The additional penalty function, although formulated with very limited experimental data, thus makes it significantly easier to find the realistic structures. 

We have demonstrated that utilizing incomplete diffraction data can significantly improve the efficiency of computational structure prediction.  
A structure that cannot be experimentally determined due to poor diffraction data can now be determined with this method.
Any experimental data can be adopted with this method to formulate the penalty function $D$; therefore, a way is opened to determine the structure of not only crystals but also other difficult to determine targets, such as surfaces, interfaces, and glass.

This work was supported by Elements Strategy Initiative to Form Core Research Center in Japan, Japan Society for the Promotion of Science KAKENHI Grant No. 17H02930 and ``Materials research by Information Integration'' Initiative (M$\rm{I^2}$I) project of the Support Program for Starting Up Innovation Hub from Japan Science and Technology Agency (JST). N.T. and D.A. were supported by the Japan Society for the Promotion of Science through the Program for Leading Graduate Schools (MERIT). The authors thank Y. Tanaka for providing the programs for X-ray diffraction patterns analysis. 

\section*{Methods}
\noindent
{\bf Details of tuning the control parameter $\bf{\alpha}$.}
The efficiency of this method depends on the value of $\alpha$. Here we estimate a suitable value. 
By adding the value of the penalty function corresponding to the melting temperature to the local minima of the energy surface, atoms are not trapped in local minima and can explore the energy surface to locate the global minimum. 
Therefore, taking the degree of freedom into account, the order of $\alpha$ is estimated as:
\eqn{
\alpha N \overline{D} \gtrsim \frac{3}{2}N k_B T_m,
}
where $N$ is the number of atoms in the simulation cell, $\overline{D}$ denotes the typical value of the penalty function $D$, $k_B$ is the Boltzmann constant and $T_m$ is the melting temperature of the system. The melting temperature for three target materials was roughly estimated as $T_m \sim 5000 \,\rm{K}$ by small-scale molecular dynamics simulations.
Based on the value of ``crystallinity'', of which the average is, for example, approximately $\frac{1}{3}$ for coesite, the typical value of $D$ can be estimated as $\frac{2}{3}$. Thus, for the target systems in this letter a suitable $\alpha$ can be roughly estimated as
 \eqn{
 \frac{4\alpha}{9 k_B} \gtrsim T_m.
 }
\begin{table*}[ht]
  	\centering
  	\begin{tabular}{cccccccc}
     	\hline \hline
   		    \cline{8-8}
	& \multicolumn{2} {c}{coesite} &  \multicolumn{2} {c}{low cristobalite} &\multicolumn{2} {c}{low quartz}  \\
    	& \, Obs.\, \cite{LL81}   & \,\,  This work \,\, & \,Obs.\cite{PDR73} \,  & \,\,This work \,\, & \, Obs.\cite{LL80} \,    & \, \, This work \,\,     \\
	\hline
	symmetery\,\, & \multicolumn{2}{c}{$C2/c$} & \multicolumn{2}{c}{$P4_12_12$}  & \multicolumn{2}{c}{$P3_221$}\\ 
   	 $a$ (\AA)  & 7.1356  & 7.23 & 4.978  &  4.99 & 4.916 & 5.02 \\
    	 $b$ (\AA) & 12.3692  & 12.74 & 4.978  &  4.99 & 4.916 & 5.02 \\
	 $c$ (\AA) & 7.1356  & 7.23 & 6.948  &  6.66 & 5.405 & 5.54 \\
	 $\alpha,\beta,\gamma$ & $\beta = 120.34^\circ$  & $120.8^\circ$  & & & $\gamma = 120.0^\circ$  & $120.0^\circ$ \\
    	\hline \hline
  	\end{tabular}
    	\caption{Structural parameters for three polymorphs of  $\rm{SiO_2}$    \label{tab:structure_parameters}} 
\end{table*}

\noindent
{\bf Details of potential energy calculation.}
The potential energy and the force are calculated using the LAMMPS\cite{SP95} package. To calculate the potential energy $E$, a pairwise interatomic potential model was adopted, which was derived from first-principles calculations\cite{ST88}. 
\\
\\
{\bf Details of techniques for optimization.} Simulated annealing for molecular dynamics was used for global optimization. The temperature of the system was controlled by the Nos\'{e}-Hoover method\cite{SN84,SN84_2,HGW85}. The integration time step was 1 fs. For simplicity, the cell parameters were fixed, although they can be optimized simultaneously in principle. To obtain the calculation results shown in Fig.~\ref{fig3}, the temperature was initially set at 5,000 K and then quenched to 0 K in the latter 3,900 steps. To obtain the calculation results shown in Fig.~\ref{fig4}, the temperature was set at 500 K for the first 4,500 steps and then quenched to 0 K for the latter 700 steps. MD visualization was created using VMD software\cite{HW96}. 
\\
\\
{\bf Details of penalty function.} 
The conventional cost function used in the structure determination from powder diffraction data is the $R$-factor\cite{KDMH96}:
\eqn{\label{R}
R = \frac{\displaystyle \sum_{\theta}|I_{\rm{obs}}(\theta) - I_{\rm{calc}}(\theta) |}{\displaystyle\sum_{\theta}|I_{\rm{obs}}(\theta)|}.
}
In contrast to the ``crystallinity'' used in our method (eq.\ref{lambda}), in the $R$-factor, all information from observed diffraction patterns is taken into account, and it is thus sensitive to background noise.

In the simulation, the X-ray wavelength was set to 1.540593 \AA, which corresponds to that for Cu K$_\alpha$ radiation. The atomic scattering factors given in international tables for crystallography\cite{EP06} were used. The diffraction angle resolution $\Delta \theta$ was set to $0.1^\circ$, which is larger than the experimental value.

\section*{Supplementary Information}
\noindent
{\bf  Structural Parameters}  Structural parameters used in this work (Supplementary Table I.) are the same as those presented in the paper\cite{ST88}, in which the potential model was proposed.




\end{document}